\documentclass[a4paper,11pt]{article}
\pdfoutput=1 

\usepackage{jinstpub} 

\usepackage{subcaption}
\usepackage{graphicx}

\title{\boldmath Development of an Analog Front-End Using NINO ASIC for Mini-ICAL}

\author[a,1]{Ravindra R Shinde,\note{Corresponding author.}}
\author[a]{Sharad R Joshi,}
\author[a,2]{Puneet Kanwar Kaur,\note{Presently working at Synopsys India Pvt. Ltd, DLF Tech Park, Plot No-01, Sector 143A, Noida, India}}
\author[a]{Gobinda Majumder,}
\author[a]{S Pethuraj,}
\author[a]{K C Ravindran,}
\author[a]{Mandar Saraf,}
\author[a]{B Satyanarayana}

\affiliation[a]{Tata Institute of Fundamental Research,\\Dr Homi Bhabha Road, Mumbai, India}

\emailAdd{rrs@tifr.res.in}

\abstract{The mini-ICAL was supposed to be a pre-prototype detector of the Iron CALorimeter (ICAL) of the India-based Neutrino Observatory (INO) project. Resistive Plate Chamber (RPC) was the active element of these detectors. As this detector operates in avalanche mode, one needs a very fast pre-amplifier for signal processing. NINO ASIC developed for the ALICE time-of-flight detector was a potential chip to be used as the analog-front end. This paper presents an 8-in-1 pre-amplifier board developed using the NINO ASIC, a single-ended to differential signal conversion using the differential driver chip manufactured by Texas Instruments, production of the boards, lessons learned during production, testing of these boards, and interface with RPCs. Finally, its performance based on strip rates and efficiency of the detector will also be discussed.}


\arxivnumber{} 

\begin{document}
\maketitle
\flushbottom

\section{Introduction}
\label{sec:intro}
\begin{figure}
     \centering
     \begin{subfigure}[b]{0.58\textwidth}
         \centering
         \includegraphics[width=\textwidth]{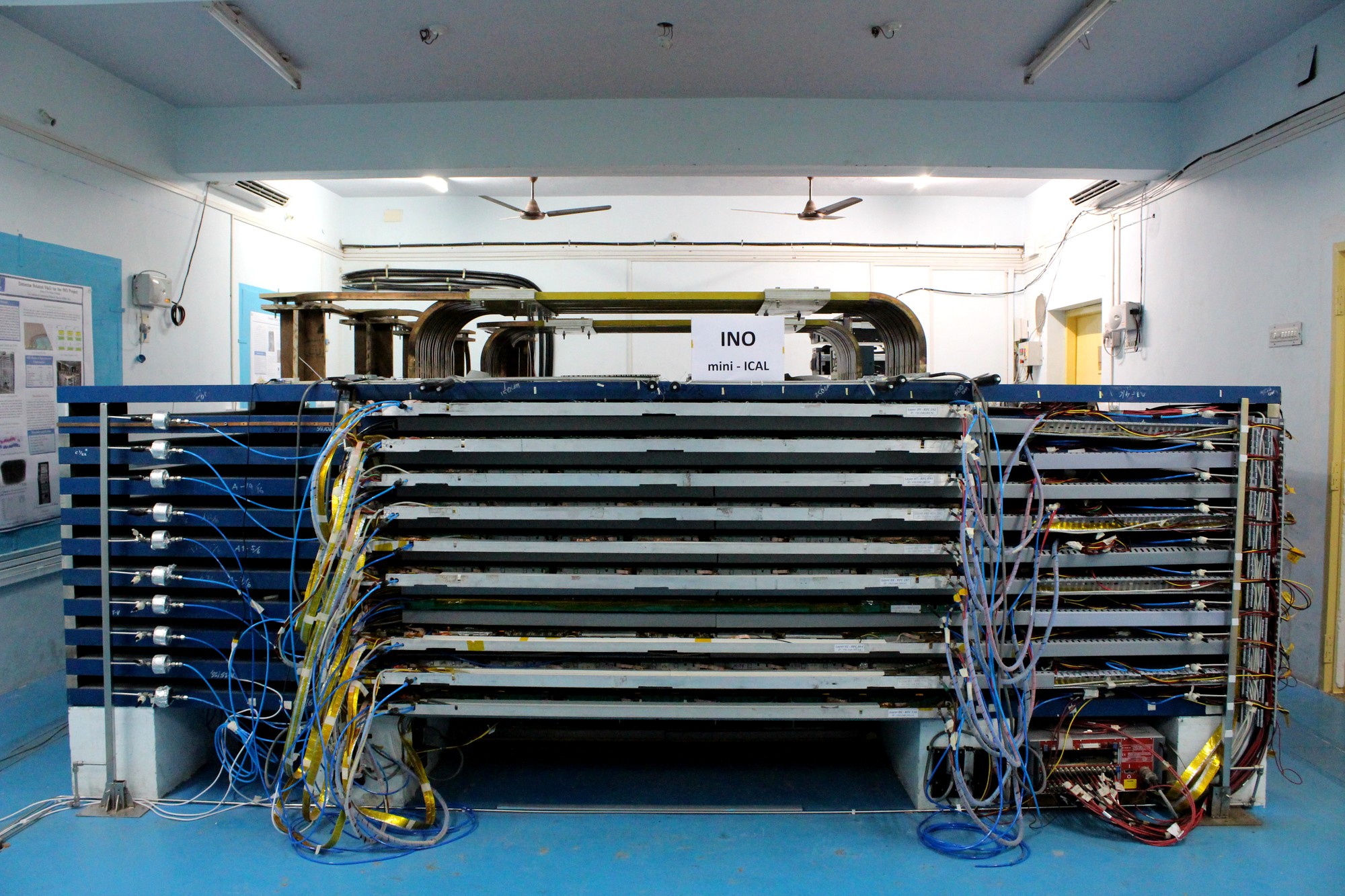}
         \caption{{\footnotesize mini-ICAL front view.}}
         \label{fig:mical:1}
     \end{subfigure}
     \hfill
     \begin{subfigure}[b]{0.40\textwidth}
         \centering
         \includegraphics[width=\textwidth]{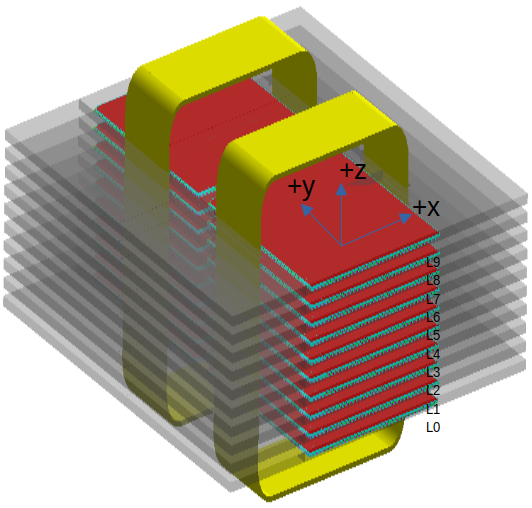}
         \caption{{\footnotesize mini-ICAL side view in GEANT4 geometry.}}
         \label{fig:mical:2}
     \end{subfigure}
        \caption{The mini-ICAL detector.}
        \label{fig:mical}
\end{figure}

The 50 kton INO-ICAL\cite{paper1} was a proposed underground high energy physics experiment at Theni, India to study neutrino oscillation parameters using atmospheric neutrinos. It was supposed to have three modules, each of size 16\,m$ \times$\,16\,m\,$\times$\,14.4\,m. Each module would have a 56\,mm thick iron layer, followed by an active element layer. The detector would have 150 layers of this combination. The Resistive Plate Chamber (RPC) was an active element in the Iron Calorimeter (ICAL) of the INO project. For the feasibility study of this mega-scale detector, a mini-ICAL detector of size 4\,m\,$\times$\,4\,m was built with 10 layers of RPCs \cite{miniICAL}. The mini-ICAL had 11 layers of 56\,mm low carbon steel with 36 copper coils to generate magnetic field of 1.5\,Tesla. Fig.~\ref{fig:mical}(a) and (b) are the detector view and simulated geometry, which gives a better view of all components of mini-ICAL. The mini-ICAL required 20 RPCs
of size 1740\,mm\,$\times$\,1845\,mm, where each layer housed 2 RPCs. An RPC is made up of 3\,mm glass electrodes with a gas gap of 2\,mm. It requires a mixture of gas, consisting of R134a (95.2\%), iso-$C_4H_{10}$ (4.5\%) and $SF_6$ (0.3\%) as ionizing medium and operates at $\sim$10\,kV and $\sim$10\,mbar pressure above the atmosphere. Top and bottom pick-up panels with strips of 30\,mm pitch are used to collect induced signals produced due to avalanche of primary ionisation on the application of high voltage. These pick-up panels are placed orthogonal to each other. In RPC operation, the positive high voltage side (defined as X-plane), produces a negative polarity signal and the negative high voltage side (defined as Y-plane), gives positive polarity signals. Therefore, two types of analog front-ends are required to accept these signals. Each RPC has 58 and 61 electronic channels for signal readout in X- and Y-plane respectively. An avalanche mode of gas operation of RPC produces pulses of small amplitude $\sim$\,1\,mV having rise time $\sim$\,1\,ns. To process such a fast pulse, one needs a very fast pre-amplifier. While the electronics team was working on ASIC development for the analog front end, thoughts were given to explore the possibility of any existing ASIC suitable for RPC operation. The NINO ASIC chip\cite{NINO} that was designed for the ALICE Time-of-Flight detector, was a potential candidate for the INO-ICAL. Thus, we have studied the ALICE 24 channel Amplifier-Discriminator board \cite{FEABoard} which has 3 NINO ASICs. NINO ASIC requires a differential signal as input. The RPC strip readout is a single ended output signal as orthogonal top and bottom panels readout separately to obtain two-dimensional positions of the traversing charge particle. Therefore, it was decided to convert the single ended signal to a differential signal to avoid attenuation of the tiny signals of the RPCs. The THS4520 \cite{THS4520} chip from Texas instruments has been used for differential signal conversion. As the NINO chip has 8 input channels, the board was designed such that it would cover 8 strips in one 8-in-1 board. In total, 320 8-in-1 NINO boards were required to collect data from the mini-ICAL detector.

\section{Prototype and final board design}
\label{sec:prototype}

\begin{figure}
\center
\includegraphics[width=0.64\linewidth]{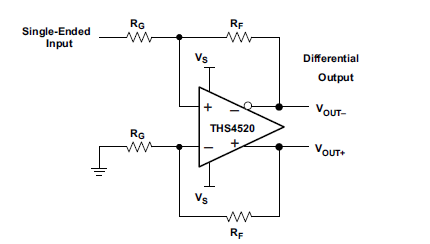}
\caption{\label{fig:fig1} Basic circuit of Differential Driver IC THS4520.}
\end{figure}

An RPC requires a very fast analog front-end to pick up fast induced pulses produced by the avalanche. To maximize the active area of the ICAL detector, there was only 25\,mm space for this front end electronics board. So we need to fit the analog front end within the space provided in the RPC tray with a height less than 25\,mm. The solution was to fix this amplifier board vertically to save space. With this constraint, an 8-in-1 pre-amplifier cum discriminator board was designed. As mentioned earlier, RPCs produce single ended signals in ICAL, so R\&D was started to convert the single ended signal to differential signals. In this process, the wide-band differential amplifier chip, THS4520 from Texas Instruments was tested.
The circuit diagram seen in Fig.~\ref{fig:fig1} shows the connections of the THS4520 to convert a single-ended signal into a differential signal with gain options of 1, 1.5 and 2. The ratio of $R_F$ to $R_G$ determines the gain of the operational amplifier. The values of $R_F$ and $R_G$ were chosen as 499\,$\Omega$ and 243\,$\Omega$ respectively to obtain the desired unity gain output. A coaxial cable is used to feed the RPC signal to the THS4520 which helps to reduce the noise in the system.

\begin{figure}
\center
\includegraphics[width=0.5\linewidth]{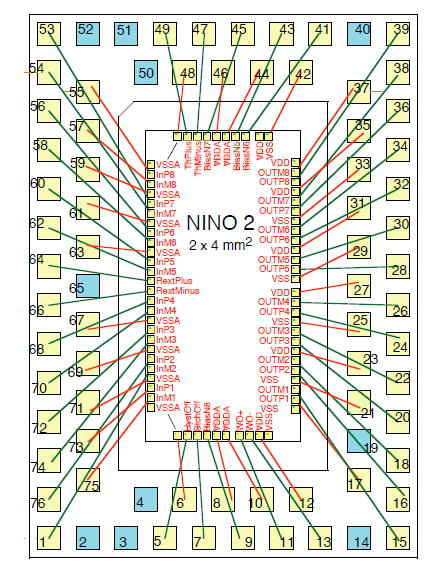}
\caption{\label{fig:fig2} NINO ASIC IC Pinout.}
\end{figure}

The differential output of the THS4520 chip is the input to the NINO ASIC. NINO is a low power, ultra-fast 8 channel front end ASIC processed in IBM 0.25\,$\mu$m CMOS technology. It requires differential input and has a threshold range between 10\,fC and 100\,fC, and discriminated LVDS output is available for further processing. It is a compact IC with 76 pins as shown in figure \ref{fig:fig2}, fitted in a 2\,mm$\times$\,4\,mm WQFN package.


The 8-in-1 pre-amplifier-cum-discriminator board has been designed with the combination of THS4520 and NINO ASIC,  in a size of 200\,mm\,$\times$\,23\,mm. In RPC operation, both positive and negative polarity signals are produced, therefore, two types of board were designed which would accept these signals. The only difference in these boards is that the output of the differential driver, which is going as input to the NINO ASIC, has been inverted. The specifications of the designed board are listed in Table~\ref{tab:tab3}.
\begin{table}[htbp]
\centering
\begin{tabular}{ll}
\hline
\textbf{Parameter} & \textbf{Value}\\ \hline
No. of analog input channels & 8 \\ 
Threshold in terms of voltage & 150\,mV to 350\,mV \\ 
No. of discriminated output channels & 8 \\ 
Nature of output signals & LVDS \\ 
Power supply & +4\,V, GND, -4\,V \\ \hline
\end{tabular}
\caption{\label{tab:tab3}Specification of the 8-in-1 NINO  board}
\end{table}

\begin{figure}
\center
\includegraphics[width=0.99\linewidth]{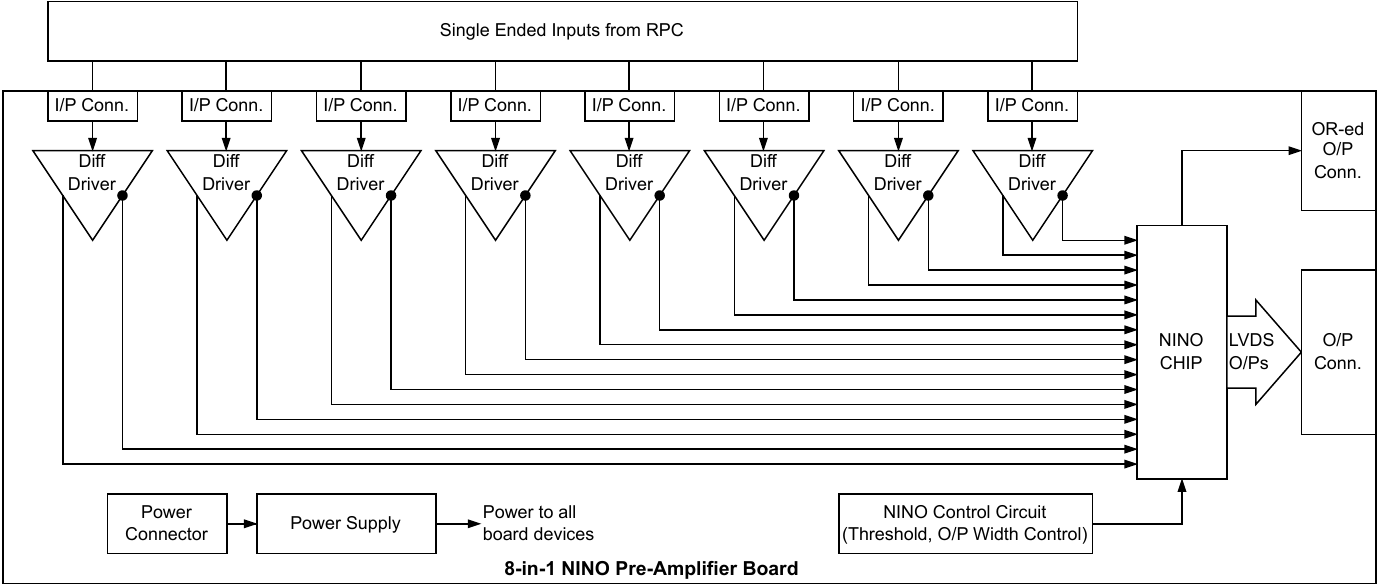}
\caption{\label{fig:fig4} Schematic of 8-in-1 NINO ASIC board.}
\end{figure}

Initially, prototype boards of X- and Y-sides were fabricated to test the performance of the circuit. The basic design blocks of the board are shown in Fig.~\ref{fig:fig4}.

\begin{figure}
\center
\includegraphics[width=0.99\linewidth]{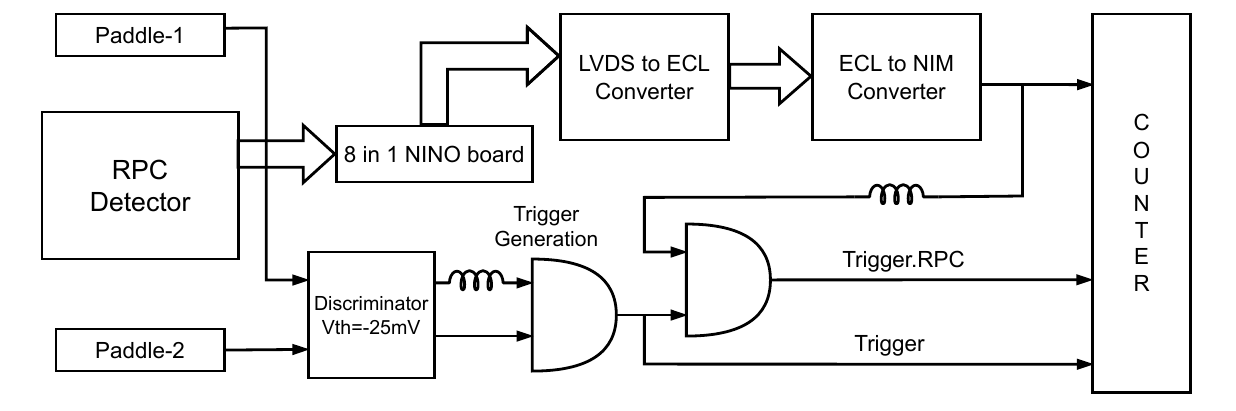}
\caption{\label{fig:nino_testing} Schematic of the test setup for NINO board testing.}
\end{figure}

Tests were carried out on boards with a pocket pulser (Phillips Scientific Model 417 NIM Pocket Pulser) and then with actual detectors. Each board was tested on the test bench with a 5\,mV attenuated pulse obtained from a NIM pocket pulsar as input. After satisfying test bench criteria for all channels, which is as follows: (a) the current drawn for positive and negative power supply lines should be $240\pm15\,mA$ and $110\pm15\,mA$ respectively; (b) discriminator threshold to vary from 150\,mV to 350\,mV; (c) noise free  pseudo LVDS output, variation of level within $\pm$100\,mV; the boards were tested with RPCs. Schematic of a cosmic ray muon telescope based test setup is shown in Fig.~\ref{fig:nino_testing}. Here, an RPC is sandwiched between two plastic scintillator paddles, Paddle-1 and Paddle-2. Trigger (T) is generated using the two paddles, which are aligned with an RPC strip under test. The LVDS output of the NINO board is converted into a NIM signal and AND-ed with the trigger, and finally connected to a counter. The ratio of (T AND RPC) to trigger gives efficiency of the RPC. The efficiencies of the RPC strips were found to be comparable to the efficiencies measured with older electronics \cite{hmcic}.

After satisfactory testing of these boards, another small lot was fabricated. The Performance of these boards was tested on the test bench and with RPCs available in the laboratory. After testing, the required number of boards were fabricated in multiple lots.

\section{Lesson learned in production of boards}
\label{sec:lesson}
\begin{figure}
\center
\includegraphics[width=0.9\linewidth]{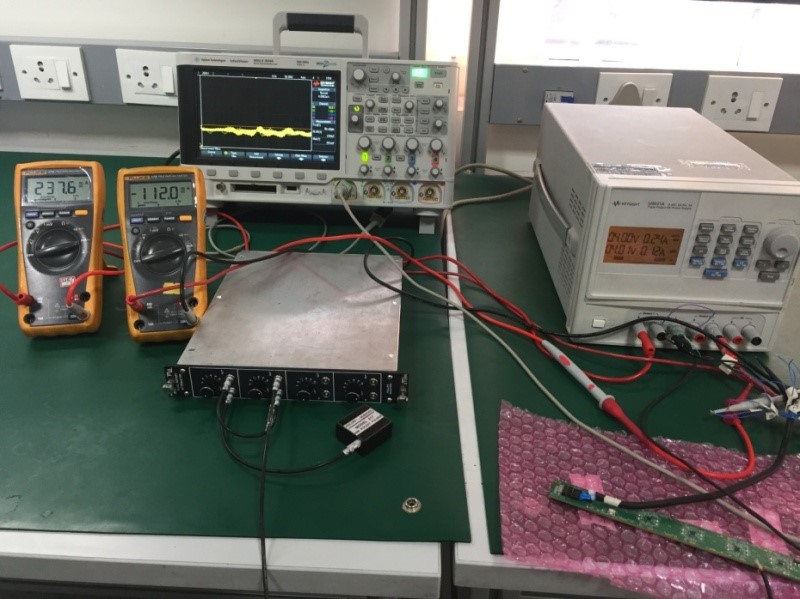}
\caption{\label{fig:fig6}Test facility at the factory.}
\end{figure}

To comply with quality control (QC), the following tasks were followed after fabrication of the boards at the factory: $(i)$ checking the resistance between power lines on the PCB; $(ii)$ checking board power supply rails; $(iii)$ checking of connectivity of NINO output pins on the board (checked by measuring voltage offset of the output pins which is caused by NINO output driver); $(iv)$ current consumption and threshold measurement; $(v)$ final verification of LVDS outputs. The specified current drawn in the board with all 8 working channels clears the quality assurance test. Fig.~\ref{fig:fig6} shows the test set-up performed in the factory. Here, the current measurement was made using digital multimeters. The combination of a NIM pocket pulser and a passive attenuator has been used to generate a 5\,mV, RPC-like pulse. The LVDS output of the respective input channel was checked on an oscilloscope.

There were several problems with the first production lot of NINO boards when lab testing was carried out. Many boards were damaged physically, including broken threshold potentiometers, dislocation of capacitors and resistors, and broken tracks. This was attributed to insufficient packaging and therefore damage during transportation. During further testing of the boards, another 22\% boards were found to be faulty. However the non-working boards had complied with the above mentioned QC checks at the factory.

\begin{figure}[htbp]
\center
\includegraphics[width=0.9\linewidth]{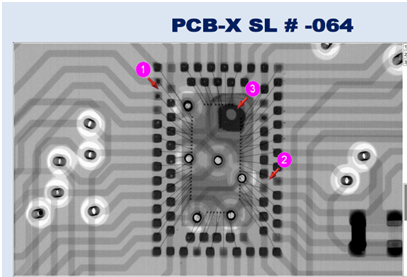}
\caption{\label{fig:fig7} X-ray scan of X-64 board having all faulty Channels.}
\end{figure}

To investigate for any soldering problems especially of the NINO chips, X-ray scanning of these boards was performed. From the X-ray results, it was clear that dry solder of the NINO pins was the root cause of the non-working boards. Due to the dry solder, even though the boards passed the factory QC tests, they failed in the lab. One of the X-ray reports is shown in Fig.~\ref{fig:fig7}. Dry solder points are seen marked with pink arrows in the board X-64, which was having all 8 non-working channels. It is seen that the VDD (power) pin is not soldered correctly, and thus no channels worked.

After reflow-soldering of the non-working boards, they were found to be working fine. Thus, it was established that dry-soldering of NINO chips was the main problem. Hence for the next lot of production, X-ray scanning was made mandatory. One more production lot of boards was fabricated and this time all boards passed quality tests without any problems, both at the factory and at the lab.

\section{Chassis development for NINO boards}
\label{sec:chasis}
\begin{figure}[htbp]
\center
\includegraphics[width=0.9\linewidth]{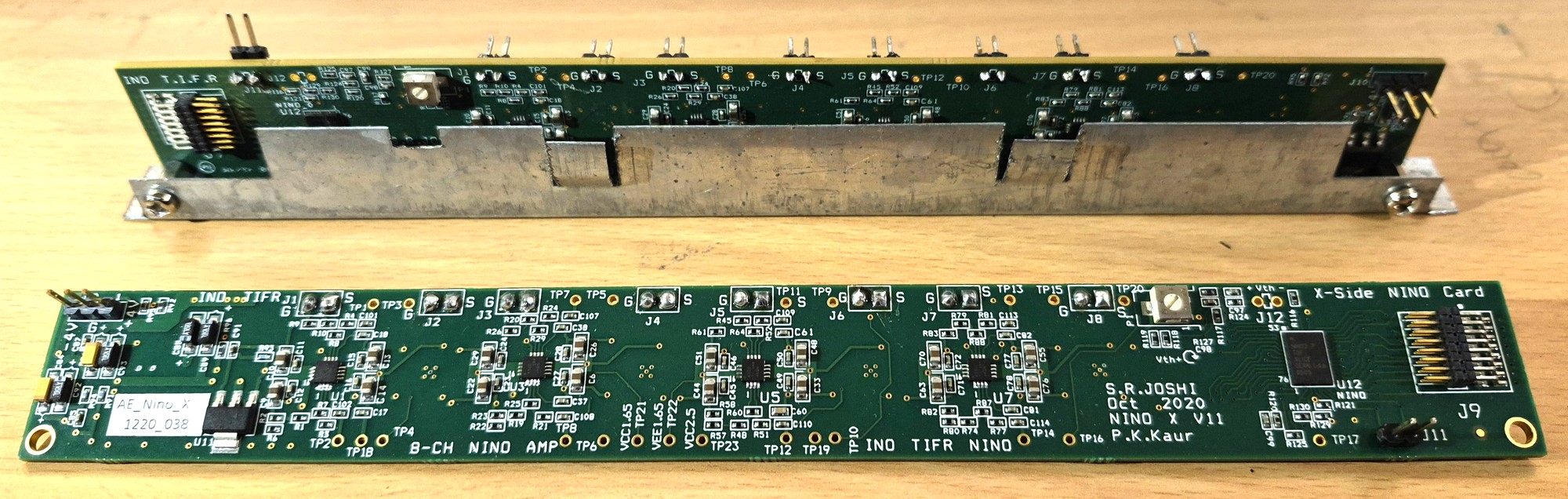}
\caption{\label{fig:fig8}NINO ASIC board with G.I. Chassis.}
\end{figure}

To avoid EMI interference and obtain noise free signal transmission, galvanised iron (G.I.) sheet chassis was designed and fabricated. These chassis were successfully integrated with the NINO boards in the RPC tray. The chassis were also helpful in mounting the NINO boards vertically in the tray. The chassis have two slots, which are useful for output signal cable routing. One of the NINO boards with chassis is shown in Fig.~\ref{fig:fig8}.

\section{Performance with RPC detector}
\label{sec:performace}
Initially, boards were tested at TIFR, Mumbai with 1\,m $\times$ 1\,m RPC and data were compared with old electronics, which was based on HMC ICs \cite{hmcic}. The data show that the efficiency and noise rate of RPC is as good as the old front end board which is listed in Table~\ref{tab:tab5}.

\begin{figure}[htbp]
\center
\includegraphics[width=0.9\linewidth]{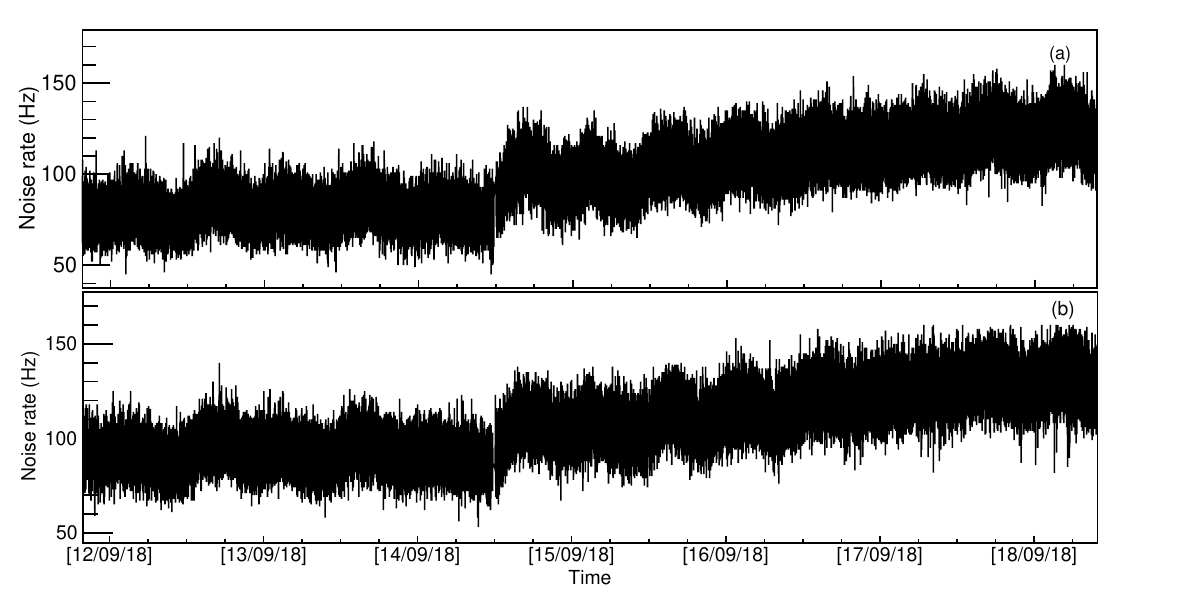}

\caption{\label{fig:fig9}The temporal variation of the strip count rates in two different channels of layer-5.}
\end{figure}

\begin{table}[htbp]
\centering
\begin{tabular}{ccccc}
\hline
\textbf{Board} & \textbf{Total Events} & \textbf{Threshold} & \textbf{Efficiency (\%)}  & \textbf{Noise Rate (Hz)}\\ \hline
HMC & 1073 & -20\,mV & 92.3 & 55 \\ 
NINO &   1634 & 80\,fC & 92.39 & 45.31\\ 
(Diff Driver Gain 1) &       &        &      & \\ \hline
\end{tabular}
\caption{\label{tab:tab5}Comparison of performance of NINO board with old HMC based analog front end.}
\end{table}
The major goal of making the NINO front-end was to study its performance using a commissioned mini-ICAL detector. The detector parameters were studied using the NINO-ASIC front-end board. In the preliminary stage, dark count rates were acquired to monitor the stability of the detector. The variation of the count rates over a week is shown in Fig.~\ref{fig:fig9} and it is found that the noise is under control. There is a 12 hour periodicity of noise rates due to the variation of atmospheric pressure. Also, the increase in noise rates from 15th September is due to a change in the fraction of SF$_6$ from 0.5\% to 0.3\% in the gas mixing unit, which slowly reduces the fraction of SF$_6$ in the chamber.

\begin{figure}[htbp]
\center
\includegraphics[width=0.99\linewidth]{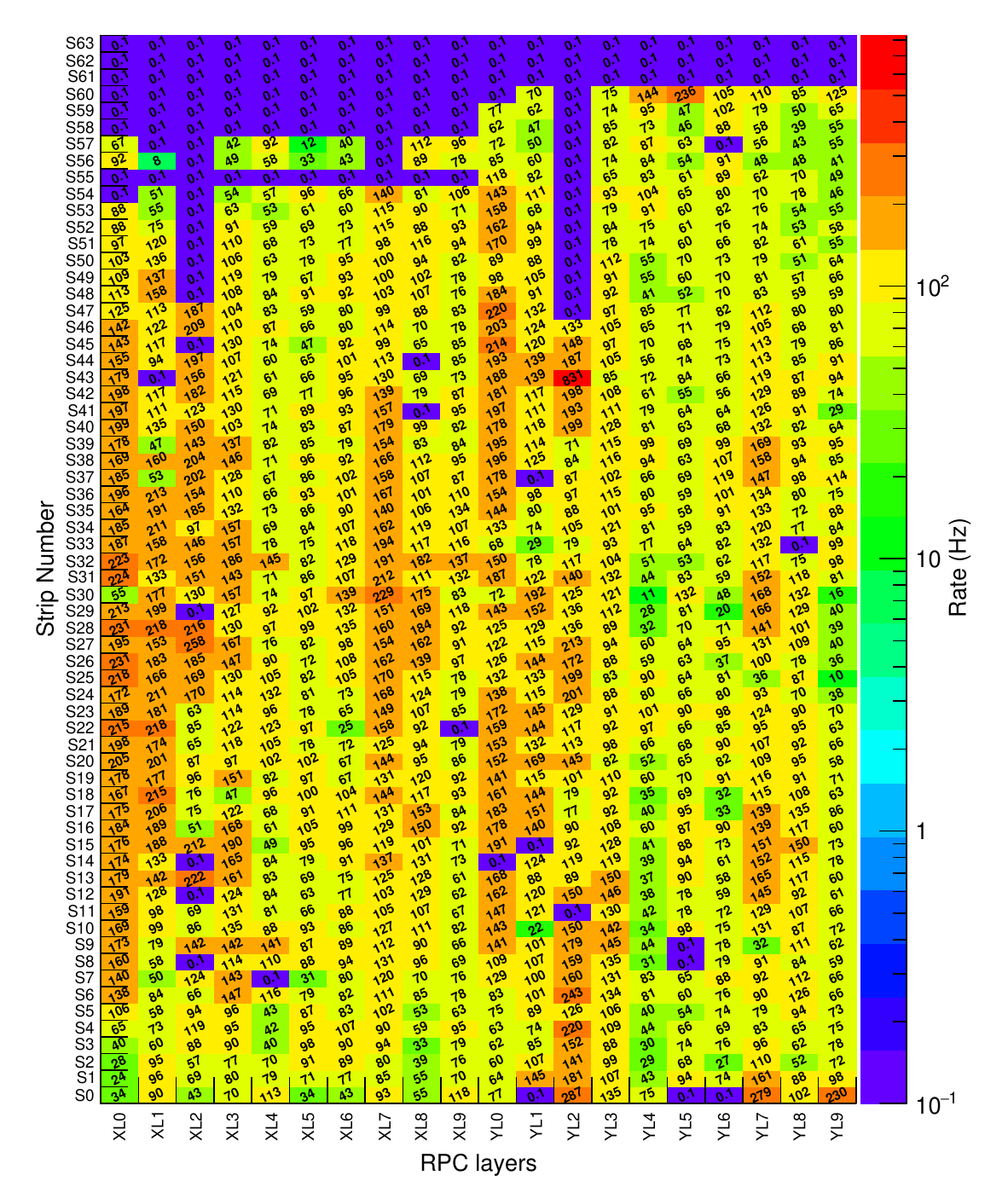}
\caption{\label{fig:noise_2D} The overall count rates of individual strips in mini-ICAL.}
\end{figure}

The necessary conditions before installation are set on the dark current of the chamber and dark count rates. The RPC currents were monitored for a few days until they reached minimum saturation (<50\,nA) and dark counts settled to less than 200\,Hz in the proper operating voltage. Using the dark count of the individual strips, the problematic front-end channels were checked if the count rate was above the normal or much below the normal rate. Fig.~\ref{fig:noise_2D}, shows the strip count rate of X- and Y-plane (in Hz). The count rates of dead strips are fixed as 0.1\,Hz for better visualization. As the number of strips in X- and Y-plane are 58 and 61, the data were obtained for all front-end boards (64 channels per plane per layer). The strip number 55 of X-plane in all the layers was electronically masked due to the pickup from the ripples in the HV connection on the RPC as the HV connection is in the vicinity of strip number 55 in X-plane.

\begin{figure}[htbp]
\center
\includegraphics[width=0.99\linewidth]{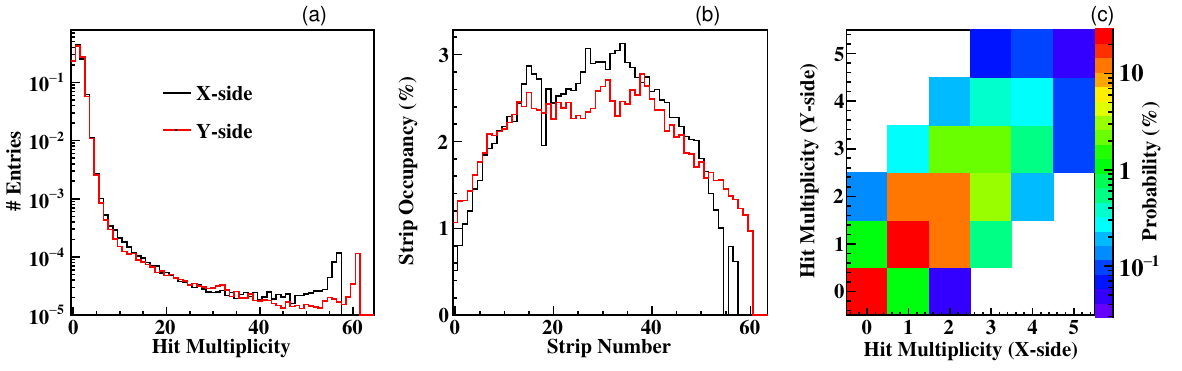}
\caption{\label{fig:fig11}(a), (b) and (c) are strip multiplicity, occupancy and correlation between the multiplicities in X- and Y-plane respectively.}
\end{figure}

After the final installation of the RPC detector assembly, including front-end and back-end, various properties of the detectors were obtained. The cosmic muon triggered event gave a better understanding of its performance. The crucial performance parameter is strip multiplicity, which is shown in Fig~\ref{fig:fig11}(a). It was estimated as the fraction of events with different multiplicities of signal with respect to the total number of triggered events. As the detector is operated in the avalanche mode, the strip multiplicity is expected to be localized mostly within three strips. Thus, the average strip multiplicity obtained with the triggered data is $\sim$1.3 strip unit. The signal multiplicity above three are primarily due to the large streamer pulses and hadron shower events in the detector. The peak in the strip multiplicity near 58 in X-plane and 61 in the Y-plane is due to the correlated electronic noise in the signal path. The strip multiplicities can be used to understand the time to time variation in the performance of the front-end boards. The occupancy (\%) of a strip represents the fraction of events registered in the strip to the total number of triggered events. Fig.~\ref{fig:fig11}(b) shows the strip occupancy in X- and Y-planes. Due to the acceptance of the trigger layers, the strip occupancy is maximum $\sim$3\% in the middle of the detector. The strip multiplicity and occupancy can be used to infer the plane dependent issues. A better understanding of the correlated behaviour in the X- and Y-planes could be obtained with the help of strip multiplicity distribution in X- and Y-planes, which is shown in Fig.~\ref{fig:fig11}(c). It is expected to have a maximum fraction in the diagonal, if there is no significant noise in the particular plane. Correlations of the strip multiplicities in X- and Y-planes are not due to the noise of individual pickup planes (NINO boards), but are due to RPC and/or external noises.

\begin{figure}[htbp]
\center
\includegraphics[width=0.99\linewidth]{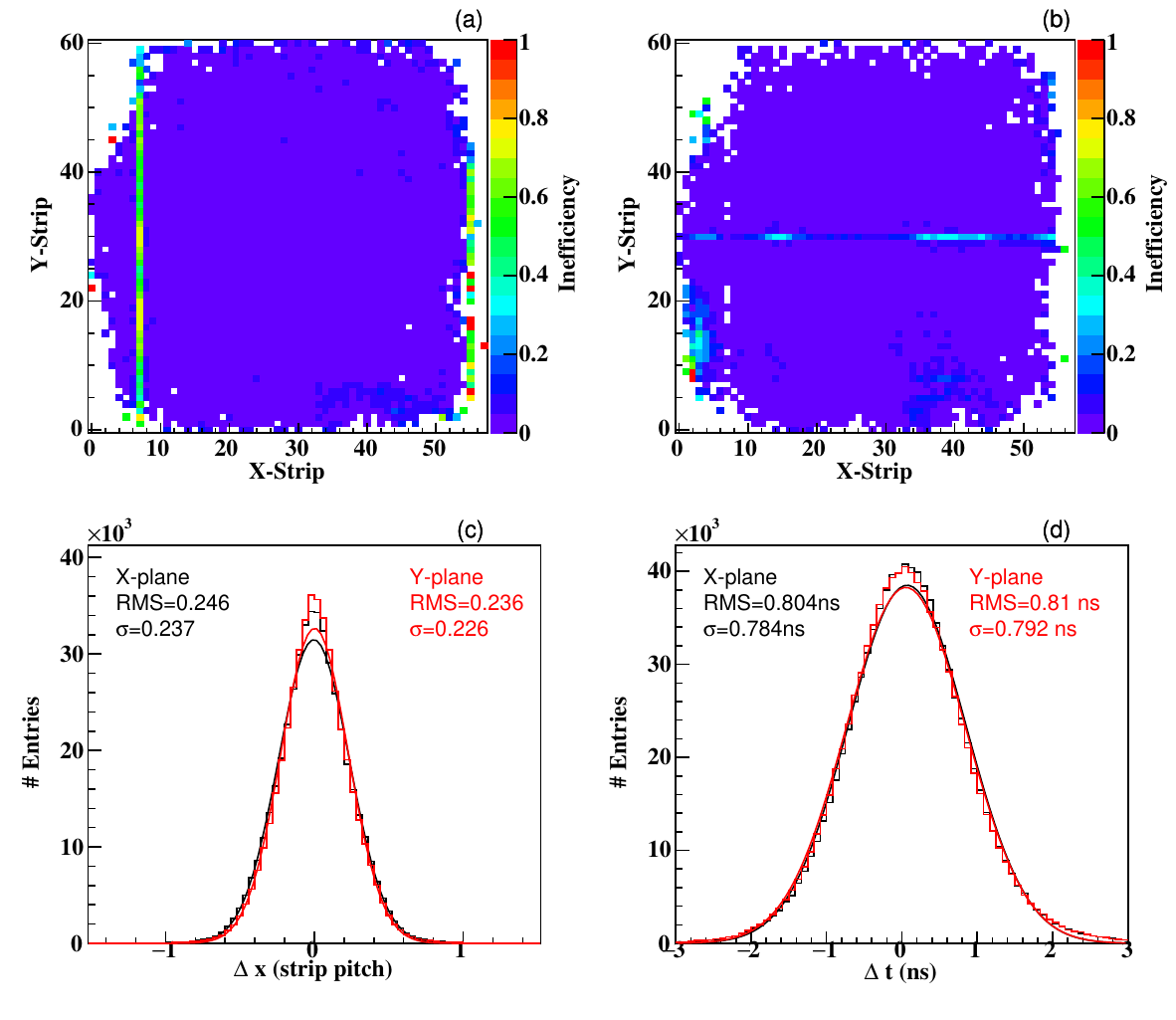}
\caption{\label{fig:fig12} The uncorrelated efficiency in layer-3, (a) for X-plane, (b) for Y-plane and (c) and (d) are the position and time resolution of the RPC detector respectively.}
\end{figure}

The issues specific to the particular region of the detector or particular plane is better visualised with the pixel wise efficiencies \cite{pethupaper}, where the size of the pixel is 3\,cm $\times$ 3\,cm matching the strip width. Pixel-wise efficiencies (or inefficiencies) are one of the opted ways to understand the behaviour of the detector along with the whole electronics chain. In this regard, the correlated and uncorrelated inefficiencies of the detector were estimated. The correlated inefficiency in the pixel is the fraction of events that do not have hits within the one strip unit in both X- and Y-planes out of the total number of traversed muons in that pixel. The uncorrelated inefficiency in the X-plane (or Y-plane) is the fraction of the events which do not have valid hits in the X-plane (or Y-plane) but have valid hits in the Y-plane (or X-plane). Fig.~\ref{fig:fig12} (a) shows the uncorrelated muon detection inefficiency in the X-side pickup panel. The white spots are due to global inefficiency of the RPC, where neither X- nor Y-planes have any signal. One band of large inefficiency shows a dead channel. Similarly, Fig.~\ref{fig:fig12} (b) shows an uncorrelated inefficiency in Y-side, which also has a dead strip. Such an effect due a dead strip or channel can be better visualised by using this uncorrelated inefficiency map. Fig.~\ref{fig:fig12} (c and d) show the position and time resolutions respectively of the RPC in both X- and Y-planes, which show better results than with the HMC preamplifier \cite{refapoorva}. The RMS and $\sigma$ in the plots implies the statistical root mean square values of the distribution and fitted Gaussian width respectively. Due to different sources of the resolutions, those are not fitted properly with a single Gaussian.

\begin{figure}[htbp]
\center
\includegraphics[width=0.9\linewidth]{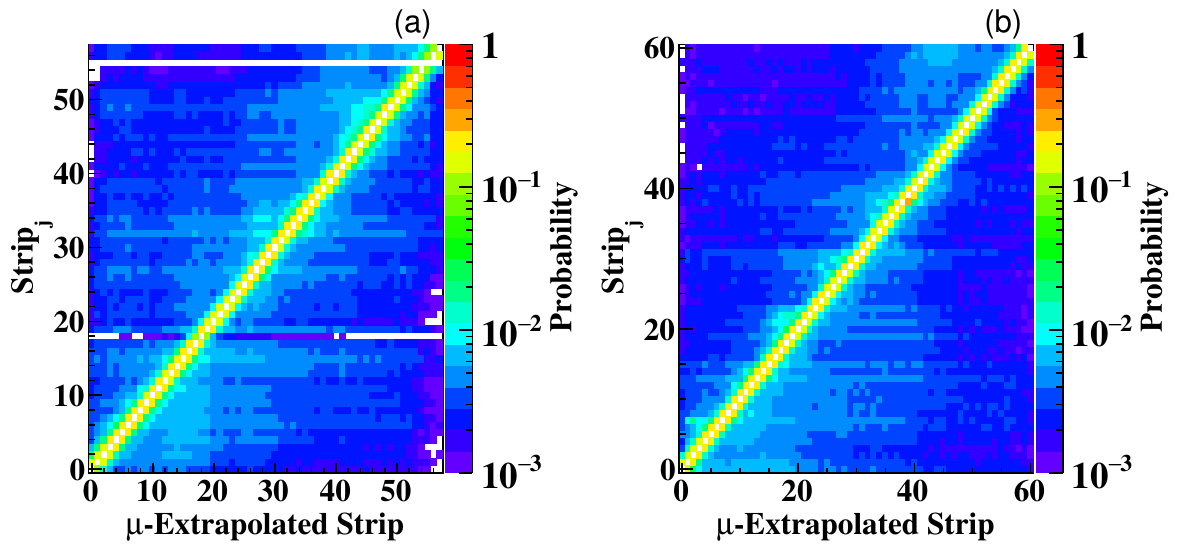}
\caption{\label{fig:fig13} (a) and (b) are the cross-talk probabilities of the muon-extrapolated strip versus any other strip in X- and Y-planes respectively.}
\end{figure}

The main purpose of choosing the NINO-ASIC based front-end is to obtain data with low noise and negligible cross-talk in the front-end boards. The cross-talk is estimated using the selected muon trajectories. The cross-talk between the strips is estimated with the correlation of hists where muon passes through one strip and any other simultaneously observed hits in the plane. Fig~\ref{fig:fig13} (a) and (b) show the cross-talk probabilities in the X- and Y-planes respectively. As anticipated, the maximum probability is observed on the diagonal, thus, the diagonal entries are removed. Also, comparatively larger correlations in nearby strips are due to the multiplicity of more than one. Each NINO boards have 8 channels and there is no discontinuity at the boundaries implies that the correlation is mainly due to the avalanches in the gas gap. The small fraction of the entries in the off diagonals is mostly due to the streamer formation in the RPC and correlated electronic noises due to external sources.

The mini-ICAL detector operated under magnet de-energised conditions was used to study the detector related parameters such as efficiency, position and time resolution of the detector. The position resolution of the experiment affects the momentum estimation of the cosmic muons in the mini-ICAL. The low noise and extremely low cross-talk probability of the NINO preamplifier, were expected to provide better position resolution. The overall position resolution of X- and Y-planes of one of the RPC detectors in mini-ICAL are 0.237 strip unit (7.11\,mm) and 0.226 strip unit (6.78\,mm) respectively. The time resolutions obtained in X- and Y-planes are 0.784\,ns and 0.792\,ns respectively.

\section{Conclusion}
\label{sec:concl}
The THS4520 differential driver chip has been tested and successfully integrated with NINO ASIC to obtain differential inputs. The 8-in-1 preamplifier cum discriminator board has been successfully designed and manufactured in required dimensions of 200\,mm\,$\times$\,23\,mm. G.I. chassis have been fabricated and used for the installation of NINO boards on RPC trays. The NINO ASIC based front-end has given low noise and negligible cross-talk, leading to better position and time resolution estimations. The mini-ICAL RPCs were operated with the NINO ASIC analog front-end and it was established that the NINO ASIC was one of the suitable options as a front-end for the ICAL detector.


\acknowledgments
The authors thank ALICE collaborators and CERN Knowledge Transfer Group for providing us with the valuable NINO ASIC chips. We acknowledge the efforts of  Ms. Darshana Gonji, Mr. Santosh Chavan , Mr. Vishal Asgolkar and Mr. Ganesh Ghodke during fabrication and testing of the boards. We are also grateful to IICHEP transit campus staff for providing excellent support during installation and maintenance.



\end{document}